\documentclass[10pt]{article}
\usepackage[article,nohylinks]{subhas}
\newcommand\WT{\func{wt}} 
\usepackage{times}

\begin{document}
\title{Some Results On Convex Greedy Embedding Conjecture for $3$-Connected Planar Graphs}
\myauthor
\maketitle              

\begin{abstract}
A greedy embedding of a graph $G = (V,E)$ into a metric space $(X,d)$ is a function $x : V(G) \rightarrow X$ such that in the embedding for every pair of non-adjacent vertices $x(s), x(t)$ there exists another vertex $x(u)$ adjacent to $x(s)$ which is closer to $x(t)$ than $x(s)$. This notion of greedy embedding was defined by Papadimitriou and Ratajczak (Theor. Comput. Sci. 2005), where authors conjectured that every $3$-connected planar graph has a greedy embedding (possibly planar and convex) in the Euclidean plane. Recently, greedy embedding conjecture has been proved by Leighton and Moitra (FOCS 2008). However, their algorithm do not result in a drawing that is planar and convex for all $3$-connected planar graph in the Euclidean plane. In this work we consider the planar convex greedy embedding conjecture and make some progress. We derive a new characterization of planar convex greedy embedding that given a $3$-connected planar graph $G = (V,E)$, an embedding $x: V \rightarrow \bbbr^2$ of $G$ is a planar convex greedy embedding if and only if, in the embedding $x$, weight of the maximum weight spanning tree ($T$) and weight of the minimum weight spanning tree ($\func{MST}$) satisfies $\WT(T)/\WT(\func{MST}) \leq \left(\card{V}-1\right)^{1 - \delta}$, for some $0 < \delta \leq 1$. In order to present this result we define a notion of weak greedy embedding. For $\beta \geq 1$ a $\beta$--weak greedy embedding of a graph is a planar embedding $x : V(G) \rightarrow X$ such that for every pair of non-adjacent vertices $x(s), x(t)$ there exists a vertex $x(u)$ adjacent to $x(s)$ such that distance between $x(u)$ and $x(t)$ is at most $\beta$ times the distance between $x(s)$ and $x(t)$. We show that any three connected planar graph $G = (V,E)$ has a $\beta$--weak greedy planar convex embedding in the Euclidean plane with $\beta \in [1,2\sqrt{2} \cdot d(G)]$, where $d(G)$ is the ratio of maximum and minimum distance between pair of vertices in the embedding of $G$. Finally, we also show that this bound is tight for well known Tutte embedding of $3$-connected planar graphs in the Euclidean plane - which is planar and convex. 
\end{abstract}
\section{Introduction}
\label{SEC1}
\subsection{Greedy embedding conjecture} 
An \emph{embedding} of an undirected graph $G = (V,E)$ in a metric space $\left(X,d\right)$ is a mapping $x : V\left(G\right) \rightarrow X$. In this work we will be concerned with a special case when $X$ is the plane ($\bbbr^2$) endowed with the Euclidean (i.e. $l_2$) metric. The function $x$ then maps each edge of the graph $G$ to the line-segments joining the images of its end points. We say that embedding is \emph{planar} when no two such line-segments (edges) intersect at any point other than their end points. Let $d\left(u,v\right)$ denote the Euclidean distance between two points $u$ and $v$. 
\begin{definition} {\bf Greedy embedding (\cite{PR05}):}
A \emph{greedy embedding} $x$ of a graph $G = (V,E)$ into a metric space $(X,d)$ is a function $x : V(G) \rightarrow X$ with the following property: for every pair of non-adjacent vertices $s,t \in V\left(G\right)$ there exists a vertex $u \in V\left(G\right)$ adjacent to $s$ such that $d\left(x\left(u\right), x\left(t\right)\right) < d\left(x\left(s\right), x\left(t\right)\right)$.
\end{definition}
This notion of greedy embedding was defined by Papadimitriou and Ratajczak in \cite{PR05}. They have presented graphs which do not admit a greedy embedding in the Euclidean plane, and conjectured following:
\begin{conjecture}[Greedy embedding conjecture]
\label{C1}
Every $3$-connected planar graph has a greedy embedding in the Euclidean plane.
\end{conjecture}
A \emph{convex embedding} of a planar graph is a ``planar embedding'' with a property that all faces, including the external faces are ``convex''. Additionally, Papadimitriou and Ratajczak stated the following stronger form of the conjecture:
\begin{conjecture}[Convex greedy embedding conjecture]
\label{C2}
Every $3$-connected planar graph has a greedy convex embedding in the Euclidean plane.
\end{conjecture}
Note that every $3$-connected planar graph has a convex embedding in the Euclidean plane (using Tutte's rubber band algorithm \cite{Tutte,LLW88,Lov86,Tho04}). In \cite{PR05} it was shown that $K_{k,5k+1}$ admits no greedy embedding for $k > 0$. Which imply that both hypotheses of the conjecture are necessary: there exist graphs that are planar but not $3$-connected ($K_{2,11}$), or $3$-connected but not planar ($K_{3,16}$), that does not admits any greedy embedding. Also, they show that high connectivity alone does not guarantee a greedy embedding. Papadimitriou and Ratajczak in \cite{PR05} also provided examples of graphs which have a greedy embedding (e.g., Hamiltonian graphs). Note that if $H \subseteq G$ is a spanning subgraph of $G$, i.e. $V(H) = V(G)$ then every greedy embedding of $H$ is also a greedy embedding of $G$. Hence, the conjecture extends to any graph having a $3$-connected planar spanning subgraph. \par
\subsection{Known results} 
Recently, greedy embedding conjecture (conjecture-\ref{C1}) has been proved in \cite{LM08}. In \cite{LM08} authors construct a greedy embedding into the Euclidean plane for all circuit graphs -- which is a generalization of $3$-connected planar graphs. Similar result was independently discovered by Angelini, Frati and Grilli \cite{AFG08}.
\begin{theorem}{\bf (\cite{LM08})}
\label{THM:LM08}
Any $3$-connected graph $G$ without having a $K_{3,3}$ minor admits a greedy embedding into the Euclidean plane.
\end{theorem}
Also, recently convex greedy embedding conjecture (conjecture-\ref{C2}) has been proved for the case of all planar triangulations \cite{Dha08} (existentially, using probabilistic methods). Note that the Delaunay triangulation of any set of points in the plane is known to be greedy \cite{BM99}, and a variant of greedy algorithm (greedy-compass algorithm) of \cite{BMBCDFML00} works for all planar triangulations.\par
Surely convex greedy embedding conjecture (conjecture-\ref{C2}) implies conjecture-\ref{C1}, however not otherwise. The greedy embedding algorithm presented in \cite{LM08,AFG08} does not necessarily produce a convex greedy embedding \cite{AM08,FF09}, and in fact the embedding may not even be a planar one. In this work we consider the convex greedy embedding conjecture (conjecture-\ref{C2}).\par
An alternative way to view the greedy embedding is to consider following path finding algorithm (see Algorithm \ref{GALGO}) on a graph $G = \left(V,E\right)$ and given embedding $x$. The algorithm in every step recursively selects a vertex that is closer to destination than current vertex. To simplify notation we write $d\left(s,t\right)$ in place of $d\left(x(s),x(t)\right)$, when embedding $x$ is given.
\begin{algorithm}
\dontprintsemicolon
\textbf{Algorithm} $\lang{GREEDY}\left(s,t \right)$\\
\eIf{$s = t$}
{
	return $\func{success}$.\;
}{
	\eIf{$\exists u \text{ adjacent to } s \text{ such that } d\left(u,t\right) < d\left(s,t\right)$}{
		$\lang{GREEDY}\left(u,t\right)$.\;
	}{
		return $\func{failure}$.\;
	}	
}
\label{GALGO}
\caption{Greedy path finding}
\end{algorithm}
Clearly, if $x$ is a greedy embedding of $G$ then for any choice of $s,t \in V$, we have a \emph{distance decreasing path} $s = v_0,v_1, \ldots,v_m = t$, such that for $i = 1, \ldots, m$, $d\left(x\left(v_i\right),x\left(v_m\right)\right) < d\left(x\left(v_{i-1}\right),x\left(v_m\right)\right)$. Thus given $G$ and $x$, a greedy path finding algorithm succeeds for every pair of vertices in $G$ iff $x$ is a greedy embedding of $G$. \par
This simple greedy path finding strategy has many useful applications in practice. Ad hoc networks and sensor nets has no universally known system of addresses like IP addresses. Also, due to resource limitations it is prohibitive to store and maintain large forwarding tables at each node in such networks. To overcome these limitations, \emph{geometric routing} uses geographic coordinates of the nodes as addresses for routing purposes \cite{KK00,KWZZ03}. Simplest of such strategy can be greedy forwarding strategy as described above (Algorithm-\ref{GALGO}). However, this simple strategy sometimes fails to deliver a packet because of the phenomenon of ``voids'' (nodes with no neighbor closer to the destination). In other words the embedding of network graph, provided by the assigned coordinates is not a greedy embedding in such cases. To address these concerns, Rao et al. \cite{RPSS03} proposed a scheme to assign coordinates using a distributed variant of Tutte embedding \cite{Tutte}. On the basis of extensive experimentation they showed that this approach makes greedy routing much more reliable. \par
%
%
%
%
Finally, Kleinberg \cite{Kle07} studied a more general but related question on this direction as: What is the least dimension of a normed vector space ${\mathcal V}$ where every graph with $n$ nodes has a greedy embedding? Kleinberg showed if ${\mathcal V}$ is a $d$-dimensional normed vector space which admits a greedy embedding of every graph with $n$ nodes, then $d = \Omega\left(\log{n}\right)$. This implies that for every finite-dimensional normed vector space ${\mathcal V}$ there exist graphs which have no greedy embedding in ${\mathcal V}$. Kleinberg also showed that there exists a finite-dimensional manifold, namely the hyperbolic plane, which admits a greedy embedding of every finite graph.\par
\subsection{Our results} 
In this work we show that given a $3$-connected planar graph $G = (V,E)$, an embedding $x: V \rightarrow \bbbr^2$ of $G$ is a planar convex greedy embedding if and only if, in the embedding $x$, weight of the maximum weight spanning tree ($\WT(T)$) and weight of the minimum weight spanning tree ($\WT(\func{MST})$) satisfies $\WT(T)/\WT(\func{MST}) \leq \left(\card{V}-1\right)^{1 - \delta}$, for some $0 < \delta \leq 1$. \par
In order to obtain this result we consider a weaker notion of greedy embedding. \emph{Weak\footnote{Not to be confused with the weaker version of the conjecture. Here weakness is w.r.t. greedy criteria, and not convexity of embedding.} greedy embedding} allows path finding algorithm to proceed as long as local optima is bounded by a factor. Formally,
\begin{definition}[Weak greedy embedding]
Let $\beta \geq 1$. A $\beta$--\emph{weak greedy embedding} $x$ of a graph $G = \left(V,E\right)$ is a planar embedding of $G$ with the following property: for every pair of non-adjacent vertices $s,t \in V\left(G\right)$ there exists a vertex $u \in V\left(G\right)$ adjacent to $s$ such that $d\left(x\left(u\right), x\left(t\right)\right) < \beta \cdot d\left(x\left(s\right), x\left(t\right)\right)$.
\end{definition}
Surely if $G$ admits a $1$-weak greedy embedding then it is greedily embeddable. We show that every $3$-connected planar graph has a $\beta$-weak greedy convex embedding in $\bbbr^2$ with $\beta \in [1,2\sqrt{2} \cdot d(G)]$, where $d(G)$ is the ratio of maximum and minimum distance between pair of vertices in the embedding of $G$. 
\subsection{Organization} Rest of the paper is organized as follows. In Section-\ref{SEC2} we present the required definitions which will be used in following sections. In section-\ref{SEC3} we define $\beta$-weak greedy convex embedding and provide a brief outline of the results. Subsequently, in section-\ref{SEC4} we derive various results on the $\beta$-weak greedy convex embedding and show that every $3$-connected planar graph has a $\beta$-weak greedy convex embedding in $\bbbr^2$ with $\beta \in [1,2\sqrt{2} \cdot d(G)]$. Finally, in section-\ref{SEC5} we derive the new condition on the weight of the minimum weight spanning tree and maximum weight spanning tree that must be satisfied in the greedy convex embedding for every $3$-connected planar graphs. Section-\ref{SEC6} contains some concluding remarks.

\section{Preliminaries}
\label{SEC2}
We will use standard graph theoretic terminology \cite{BM08}. Let $G = \left(V,E\right)$ be an undirected graph with vertex set $V$ and edge set $E$, where $\card{V} = n$. Given a set of edges $X \subseteq E\left(G\right)$, let $G\left[X\right]$ denote the subgraph of $G$ induced by $X$. For a vertex $u \in V$, let ${\mathcal N}(u) = \set{v : uv \in E}$ denote its neighborhood. A connected acyclic subgraph $T$ of $G$ is a \emph{tree}. If $V(T) = V(G)$, then $T$ is a \emph{spanning tree}. For $x,y \in V(G)$, $xy$--paths $P$ and $Q$ in $G$ are \emph{internally disjoint} if $V(P) \cap V(Q) = \set{x,y}$. Let $p(x,y)$ denote the maximum number of pair-wise internally disjoint paths between $x,y \in V(G)$. A nontrivial graph $G$ is $k$-connected if $p(u,v) \geq k$ for any two distinct vertices $u, v \in V(G)$. The \emph{connectivity} $\kappa(G)$ of $G$ is the maximum value of $k$ for which $G$ is $k$-connected.\par
%
%
\section{Weak greedy embedding of 3-connected planar graphs}
\label{SEC3}
In this section we define $\beta$-weak greedy convex embedding, and provide an outline of the proof. In rest of the section $x : V\left(G\right) \rightarrow \bbbr^2$ be a planar convex embedding of $G = (V,E)$ which produces a one-to-one mapping from $V$ to $\bbbr^2$. We shall specifically consider Tutte embedding (\cite{Tutte,LLW88,Lov86,Tho04}) and a brief description of Tutte embedding has been provided in Appendix-\ref{APX2}. Since $x$ is fixed, given a graph $G$, we will not differentiate between $v \in V(G)$ and its planar convex embedding under $x$ viz. $x(v)$.\par
First let us consider following recursive procedure for $\beta$--weak greedy path finding given in Algorithm-\ref{BWGALGO}.
\begin{algorithm}[htbp]
\dontprintsemicolon
\textbf{Algorithm} $\lang{WEAK-GREEDY}\left(s,t,\beta \right)$\\
\eIf{$s = t$}
{
	return $\func{success}$.\;
}{
	$B \define \set{v : (s,v) \in E \text{ and } d(v,t) < \beta \cdot d(s,t)}$.\;
	\eIf{$B = \emptyset$}{
		return $\func{failure}$.\;
	}{
		$\forall v \in B$: $\lang{WEAK-GREEDY}\left(v,t,\beta \right)$.\;
	}	
}
\label{BWGALGO}
\caption{$\beta$--weak greedy path finding}
\end{algorithm}
If $\beta$ is chosen as the minimum value such that $\forall t \in V - \set{s}$ at least one branch of this recursive procedure returns $\func{success}$ then we will call that value of $\beta = \beta_s$ optimal for vertex $s$. Given $(s,\beta_s)$ for a vertex $t \in V - \set{s}$ there can be more than one $\beta_s$--weak greedy path from $s$ to $t$. Let $H(s,\beta_s) \subseteq G$ be a subgraph of $G$ induced by all vertices and edges of $\beta_s$--weak greedy $st$--paths for all possible terminal vertex $t \in V - \set{s}$. Let $T(s,\beta_s)$ be any spanning tree of $H(s,\beta_s)$. Surely, $T(s,\beta_s)$ has unique $\beta_s$--weak greedy $st$--paths for all possible terminal vertex $t \in V - \set{s}$ from $s$. We will call $T_s = T(s,\beta_s)$ optimal weak greedy tree w.r.t vertex $s$. Define $\beta_{\max} \define \max_{s \in V}{\set{\beta_s}}$. We note that procedure $\lang{WEAK-GREEDY}\left(s,t,\beta_{\max} \right)$ with parameter $\beta_{\max}$ succeeds to find at least one $\beta_{\max}$--weak greedy $st$--paths for all possible vertex pairs $s,t \in V$. In following our objective will be to obtain a bound on $\beta_{\max}$ for any $3$-connected planar graph $G$ under embedding $x$. To obtain this bound we will use the properties of weak greedy trees.\par
What follows is a brief description of how we obtain the stated results. In the planar convex embedding of $G$, let weight of an edge $e = uv$ be its length i.e. $\WT(e) = d(u,v)$. Define $\WT(T(s,\beta_s)) = \sum_{e \in E(T(s,\beta_s))}{\WT(e)}$. We obtain a lower and upper bound on the weight of $T(s,\beta_s)$. On the other hand we also obtain a upper bound on the weight of any spanning tree $T$ of $G$ in its embedding $\WT(T)$, and a lower bound on the weight of any minimum spanning tree $\func{MST}$ of $G$, $\WT(\func{MST})$. Surely $\WT(\func{MST}) \leq \WT(T_s) \leq \WT(T)$, and from this we derive an upper and a lower bound on $\beta_{\max}$. Let $d_{\max}(G) = \max_{u,v \in V}{d(u,v)}$ be the diameter of $G$, and let minimum edge length in embedding of $G$ be $d_{\min}(G)$. In following (in Section-\ref{SEC41}) we derive that, 
\begin{equation}
\WT(T) \leq \sqrt{2} \cdot (\card{V} - 1) \cdot d_{\max}(G).\nonumber
\end{equation}
Subsequently (in Section-\ref{SEC42}), we show that,
\begin{equation}
d_{\max}(G) \leq \WT(\func{MST}) \leq 2.5 \cdot d^2_{\max}(G).\nonumber
\end{equation}
Finally (in Section-\ref{SEC43}), we derive upper and lower bounds on the the weight of $T(s,\beta_s)$ as:
\begin{equation}
d_{\min}(G) \cdot \left(\beta_{\max} - 1\right) \cdot \left(\card{V}-1\right) \leq \WT(T_s) \leq 2 \cdot d_{\max}(G) \cdot \left(\frac{\beta_{max}^{\card{V}-1} - 1}{\beta_{max}-1}\right)\nonumber
\end{equation}
Using the fact that $\WT(\func{MST}) \leq \WT(T_s) \leq \WT(T)$, we than show using the bounds described above - that any three connected planar graph has a $\beta$-weak greedy convex embedding in $\bbbr^2$ with $\beta \in [1,2\sqrt{2} \cdot d(G)]$, where $d(G) = d_{\max}(G)/d_{\min}(G)$. Our main result states that given a $3$-connected planar graph $G = (V,E)$, an embedding $x: V \rightarrow \bbbr^2$ of $G$ is a planar convex greedy embedding if and only if, in the embedding $x$, weight of the maximum weight spanning tree ($\WT(T)$) and weight of the minimum weight spanning tree ($\WT(\func{MST})$) satisfies $\WT(T)/\WT(\func{MST}) \leq \left(\card{V}-1\right)^{1 - \delta}$, for some $0 < \delta \leq 1$. To establish one side of this implication we use the bounds on the weight of $T(s,\beta_s)$ and the upper bound on the weight of the $\func{MST}$. 
%
%
\section{Bounding the weight of trees}
\label{SEC4}
In following we first describe upper bound on the weight of any spanning tree $T$ of $G$ in its planar convex embedding. In order to obtain this bound we use some ideas from \cite{MS92}.
\subsection{Upper bound on the weight of spanning tree}
\label{SEC41}
Given a graph $G = (V,E)$ and its planar convex embedding, let $d_{\max}(G) = \max_{u,v \in V}{d(u,v)}$ be the diameter of $G$ and let $T$ be any spanning tree of $G$. For $i = 1, \ldots, \card{V}-1$ let $e_i$ be $i$th edge of $T$ (for a fixed indexing of edges). Let $D_i$ be the open disk with center $c_i$ such that $c_i$ is the mid point of $e_i = uv$, and $D_i$ having diameter $d(u,v)$. We will call $D_i$ a diametral circle of $e_i$. Let $\bar{D_i}$ be the smallest disk (closed) that contains $D_i$. Define $D = \cup_{e_i \in E(T)}{\bar{D_i}}$. We have following claim:
\begin{lemma}
\label{L1}
$D$ is contained into a closed disk $D'$ having its center coinciding with $D$ and having diameter at most $\sqrt{2}\cdot d_{\max}(G)$.
\end{lemma}
\begin{proof}
Let $D = \cup_{e_i \in E(T)}{\bar{D_i}}$ having its center at point $c \in \bbbr^2$. Let $e = uv \in T$ be an edge -- surely $u$ and $v$ are points inside $D$. Consider the closed disk $\bar{D_{uv}}$ centered at the midpoint of $e$ having diameter $d(u,v)$. Let $c'$ be its center. Since $D'$ must contain $\bar{D_{uv}}$, worst case is when both $u$ and $v$ are at the boundary of $D$ (see Figure-\ref{fig2A}). Now let $z$ be any point on the boundary of $\bar{D_{uv}}$. We have:
\begin{figure}[htbp]
\centering
\includegraphics[viewport=80 80 270 270,width=0.35\textwidth,clip]{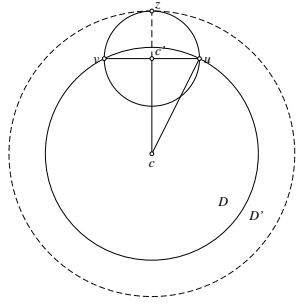}
\caption{Illustration to the proof of Lemma-\ref{L1}}
\label{fig2A}
\end{figure}
\begin{equation}
d(c,z) \leq d(c,c') + d(c',z) = d(c,c') + \frac{d(u,v)}{2}\nonumber
\end{equation}
Since, $\angle c c' u = \pi/2$,
\begin{equation}
d(c,c')^2 + d(c',u)^2 = d(c,u)^2 = \left(\frac{d_{\max}(G)}{2}\right)^2\nonumber
\end{equation}
On the other hand $d(c',u) = \frac{d(u,v)}{2}$. Hence,
\begin{equation}
d(c,z) \leq \sqrt{\frac{d^2_{\max}(G) - d(u,v)^2}{4}} + \frac{d(u,v)}{2}\nonumber
\end{equation}
Right side is maximized when $d(u,v) = d_{\max}(G)/\sqrt{2}$, and in that case $d(c,z) \leq d_{\max}(G)/\sqrt{2}$. \qed
\end{proof}
Using Lemma-\ref{L1} we can now obtain a bound on $\WT(T)$. Let $\func{Circ}(D_i)$ denote the circumference of circle $D_i$, i.e. $\func{Circ}(D_i) = \pi \cdot \WT(e_i)$.
\begin{lemma}
\label{L3}
$\WT(T) \leq \sqrt{2} \cdot (\card{V} - 1) \cdot d_{\max}(G)$
\end{lemma}
\begin{proof}
\begin{eqnarray}
\WT(T) &=& \sum\limits_{e_i \in E(T)}{\WT(e_i)} = {\frac{1}{\pi} \cdot \sum\limits_{e_i \in E(T)}{\func{Circ}(D_i)}}.\nonumber
\end{eqnarray}
Let $D'$ be a closed disk in which $D = \cup_{e_i \in E(T)}{\bar{D_i}}$ is contained, where $\bar{D_i}$ is the smallest disk (closed) that contains $D_i$. Using Lemma-\ref{L1}, and using the fact that $T$ is a spanning tree and hence have $(\card{V} - 1)$ edges, we have:
\begin{eqnarray}
\WT(T) &\leq& {\frac{1}{\pi} \cdot (\card{V} - 1) \cdot \func{Circ}(D')} \nonumber\\
&\leq& {\frac{1}{\pi} \cdot (\card{V} - 1) \cdot \left(\pi \sqrt{2} \cdot d_{\max}(G)\right)} \leq \sqrt{2} \cdot (\card{V} - 1) \cdot d_{\max}(G).\nonumber
\end{eqnarray}
\qed
\end{proof}
%
%
\subsection{Bound on the weight of minimum weight spanning tree}
\label{SEC42}
In the planar convex embedding of $G$ let $\func{MST}$ be a minimum weight spanning tree of $G$ and let $\WT(\func{MST})$ be its weight. In this section we obtain an upper and a lower bound on $\WT(\func{MST})$. Let $V \subset \bbbr^2$ be the point set given (as images of vertex set) by the embedding. Let ${\mathcal E}$ be the set of all line-segments $uv$ corresponding to the all distinct pair of end-points $u,v \in V$. Also, let $\func{EMST}$ be a spanning tree of $V$ whose edges are subset of ${\mathcal E}$ such that weight $\WT(\func{EMST})$ is minimum ($\func{EMST}$ is a Euclidean minimum spanning tree of the point set $V$). Surely, $\WT(\func{EMST}) \leq \WT(\func{MST})$: convex embedding produces a straight-line embedding of $G$, and hence the line segments corresponding to the edges of $G$ in embedding are also subset of ${\mathcal E}$. Let $u$ and $v$ be vertices having distance $d_{\max}(G)$. Any $\func{EMST}$ would connect $u$ and $v$. Hence we have:
\begin{lemma}
\label{L8}
In planar convex embedding of $G$, 
\begin{equation}
\WT(\func{MST}) \geq \WT(\func{EMST}) \geq d_{\max}(G).\nonumber
\end{equation}
\end{lemma}
We will also require upper bound on the weight of minimum spanning tree for which we have:
\begin{lemma}
\label{LMSTUB}
In planar convex embedding of $G$, 
\begin{equation}
\WT(\func{MST}) \leq \frac{5}{2} \cdot d^2_{\max}(G).\nonumber
\end{equation}
\end{lemma}
\begin{proof}
Given a graph $G = (V,E)$ and its planar convex embedding, let $d_{\max}(G) = \max_{u,v \in V}{d(u,v)}$ be the diameter of $G$ and let $\func{MST}$ be any minimum weight spanning tree of $G$. For $i = 1, \ldots, \card{V}-1$ let $e_i$ be $i$th edge of $\func{MST}$ (for a fixed indexing of edges). Let $D_i$ be the open disk with center $c_i$ such that $c_i$ is the mid point of $e_i = uv$, and $D_i$ having diameter $d(u,v)$. We will call $D_i$ a diametral circle of $e_i$. Let $\bar{D_i}$ be the smallest disk (closed) that contains $D_i$. Define $D = \cup_{e_i \in E(\func{MST})}{\bar{D_i}}$. Recall, using Lemma-\ref{L1} we have that $D$ is contained into a closed disk $D'$ having its center coinciding with $D$ and having diameter at most $\sqrt{2}\cdot d_{\max}(G)$. Let $\func{Circ}(D)$ denote the circumference of circle $D$, i.e. $\func{Circ}(D) = \pi \cdot \WT(e)$, where $D$ is a diametral circle of edge $e$. Also, let $\func{Area}(D)$ denote the area of circle $D$, i.e. $\func{Area}(D) = \pi \cdot (d/2)^2$, where $D$ is a circle having diameter $d$. Now, 
\begin{eqnarray}
\WT(\func{MST}) &=& \sum\limits_{e_i \in E(\func{MST})}{\WT(e_i)} = {\frac{1}{\pi} \cdot \sum\limits_{e_i \in E(\func{MST})}{\func{Circ}(D_i)}}.\nonumber
\end{eqnarray}
Now by Lemma-\ref{L1}, all the points that we would like to count in $\sum_{e_i \in E(\func{MST})}{\func{Circ}(D_i)}$ are contained in ${\func{Area}(D')}$. Except that some of the points that appear on the circumference of more than one circles, must be counted multiple times. In order to bound that we shall use following result from \cite{CCPRV01}. 
\begin{lemma}[Lemma-2 from \cite{CCPRV01}]
\label{LEMC}
For any point $p \in \bbbr^2$, $p$ is contained in at most five diametral circles drawn on the edges of the $\func{MST}$ of a point set $V \subset \bbbr^2$.
\end{lemma}
Using Lemma-\ref{L1}, and using the Lemma-\ref{LEMC}, we have:
\begin{eqnarray}
\WT(\func{MST}) &=& {\frac{1}{\pi} \cdot \sum\limits_{e_i \in E(\func{MST})}{\func{Circ}(D_i)}}\nonumber\\
&\leq& \frac{1}{\pi} \cdot 5 \cdot {\func{Area}(D')} \leq \frac{1}{\pi} \cdot 5 \cdot \pi \left(\frac{\sqrt{2}\cdot d_{\max}(G)}{2}\right)^2 = \frac{5}{2} \cdot d^2_{\max}(G).\nonumber
\end{eqnarray}
\qed
\end{proof}
\subsection{Bound on the weight of weak greedy trees}
\label{SEC43}
Given a graph $G = (V,E)$ and its planar convex embedding, let $T_s = T(s,\beta_s)$ be an optimal weak greedy tree w.r.t a vertex $s \in V$. Let $t$ be any leaf vertex of $T_s$, and consider the $\beta_s$--weak greedy $st$--path.
\begin{definition}[Increasing and decreasing sequence]
\label{D1}
Given a graph $G = (V,E)$ and its planar convex embedding, for $\beta_s$--weak greedy $st$--path $P_{st} = \set{s = u_0, u_1, \ldots, u_k = t}$, an ordered sequence of vertices $\set{u_{i_0}, \ldots, u_{i_r}}$ of $P_{st}$ is an increasing sequence of length $r$ if $d(u_{i_0},t) \leq \ldots \leq d( u_{i_r},t)$ holds. Similarly, an ordered sequence of vertices $\set{u_{i_0}, \ldots, u_{i_r}}$ of $P_{st}$ is a decreasing sequence of length $r$ if $d(u_{i_0},t) \geq \ldots \geq d( u_{i_r},t)$ holds. Usually, we will refer any maximal (by property of monotonically non-decreasing or non-increasing) sequence of vertices as increasing or decreasing sequence.
\end{definition}
It is straightforward to observe that if an $st$--path is $\beta_s$--weak greedy for $\beta_s > 1$, then it has a monotonically non-decreasing sequence of vertices. However, every $st$--path must have a trailing monotonically decreasing sequence that reaches $t$ (e.g. see Figure-\ref{fig1D}). We will call an increasing sequence $\set{u_{i_0}, \ldots, u_{i_r}}$ of $P_{st}$ a $\beta$-increasing sequence of length $r$ if it is maximal and for $j = 1, \ldots, r, d(u_{i_j},t) \leq \beta d(u_{i_{j-1}},t)$ holds (with equality for at least one $j$). We will denote it as $\func{inc(r,d,\beta)}$, where $d$ indicates $d(u_{i_0},t) = d$. 
\begin{lemma}
\label{L4}
Let $\func{inc(k,d,\beta)} = \set{u_{i_0}, \ldots, u_{i_k}}$ be a $\beta$-increasing sequence of length $k$ from a $\beta_s$--weak greedy $st$--path such that $d(u_{i_0},t) = d$. Then 
\begin{equation}
d (\beta^k - 1) \leq \WT(\func{inc(k,d,\beta)}) \leq d (\beta^k - 1) \left(\frac{\beta + 1}{\beta - 1}\right)\nonumber
\end{equation}
Where $\WT(\func{inc(k,d,\beta)})$ is the sum of the weight of the edges of $\func{inc(k,d,\beta)}$.
\end{lemma}
\begin{proof}
First let us bound the length $x$ of $i$th segment in $\func{inc(k,d,\beta)}$(see Figure-\ref{fig3A}). We have $d(u_{i-1},t) \leq d \beta^{i-1}$, and $d(u_{i},t) \leq d \beta^{i}$. Let $\angle{u_{i-1} t u_{i}} = \alpha$. We have $y = d \beta^{i} \sin{\alpha}$ and $z = d \beta^{i-1}( \beta \cos{\alpha} - 1)$. Since $\angle{u_{i} t' t} = \pi/2$
\begin{figure}[htbp]
\centering
\includegraphics[viewport=0 0 130 140,width=0.25\textwidth,clip]{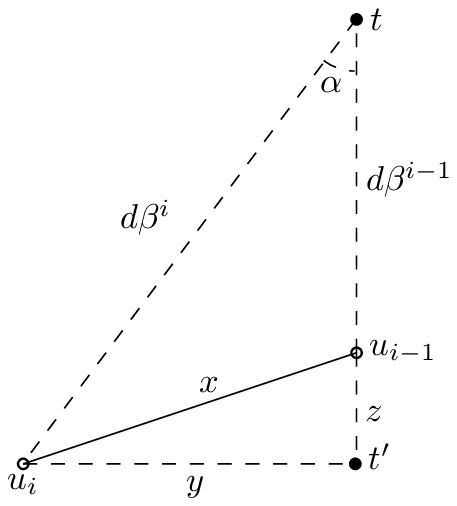}
\caption{Illustration to the proof of Lemma-\ref{L4}}
\label{fig3A}
\end{figure}
\begin{align}
x^2 = y^2 + z^2 &= (d \beta^{i-1})^2 (\beta^2 \sin^2{\alpha} + \beta^2 \cos^2{\alpha} - 2 \beta \cos{\alpha} + 1) \notag\\
&= (d \beta^{i-1})^2 (\beta^2 - 2 \beta \cos{\alpha} + 1) \notag\\
&\leq (d \beta^{i-1})^2 (\beta^2 + 2 \beta + 1) = (d \beta^{i-1})^2 (\beta + 1)^2\notag
\end{align}
So $x \leq d \beta^{i-1} (\beta + 1)$. Similarly, 
\begin{eqnarray}
x^2 &=& (d \beta^{i-1})^2 (\beta^2 - 2 \beta \cos{\alpha} + 1) \geq (d \beta^{i-1})^2 (\beta^2 - 2 \beta + 1) = (d \beta^{i-1})^2 (\beta - 1)^2\nonumber
\end{eqnarray}
Hence, $x \geq d \beta^{i-1} (\beta - 1)$. So starting at a distance $d$ from $t$ and summing over $k$ length sequence, we have for upper bound on $\WT(\func{inc(k,d,\beta)})$:
\begin{equation}
\WT(\func{inc(k,d,\beta)}) = \sum\limits_{j = 1}^{k}{d(u_{j-1},u_{j})} \leq d (\beta + 1) \sum\limits_{j = 1}^{k}{\beta^{j-1}} = d (\beta + 1) \left(\frac{\beta^k - 1}{\beta - 1}\right)\nonumber
\end{equation}
And for lower bound on $\WT(\func{inc(k,d,\beta)})$ we have,
\begin{eqnarray}
\WT(\func{inc(k,d,\beta)}) &=& \sum\limits_{j = 1}^{k}{d(u_{j-1},u_{j})} \nonumber\\
&\geq& d (\beta - 1) \sum\limits_{j = 1}^{k}{\beta^{j-1}} = d (\beta - 1) \left(\frac{\beta^k - 1}{\beta - 1}\right) = d (\beta^k - 1)\nonumber
\end{eqnarray}
\qed
\end{proof}
Like $\func{inc(r,d,\beta)}$, for $\gamma > 1$ by $\func{dec(r,d,\gamma)}$ we will denote a decreasing sequence $\set{u_{i_0}, \ldots, u_{i_r}}$ of $P_{st}$ as a $\gamma$-decreasing sequence of length $r$ if it is maximal and for $j = 1, \ldots, r$, $d(u_{i_{j-1}},t) \leq \gamma d(u_{i_j},t)$ holds (with equality for at least one $j$), where $d$ indicates $d(u_{i_0},t) = d$. 
\begin{lemma}
\label{L5}
Let $\func{dec(k,d,\gamma)} = \set{u_{i_0}, \ldots, u_{i_k}}$ be a $\gamma$-decreasing sequence of length $k$ such that $d(u_{i_0},t) = d$. Then 
\begin{equation}
d(1- \frac{1}{\gamma}) \leq \WT(\func{dec(k,d,\gamma)}) \leq d k (1+ \frac{1}{\gamma}) \nonumber
\end{equation}
\end{lemma}
\begin{proof}
A similar calculation as in the proof of Lemma-\ref{L4} shows that the length $x$ of $i$th segment is bounded from above by $(d/\gamma^{i-1})(1 + 1/\gamma)$, and from below by $(d/\gamma^{i-1})(1 - 1/\gamma)$. So starting at a distance $d$ from $t$ and summing over $k$ length sequence, we have upper bound on $\WT(\func{dec(k,d,\beta)})$:
\begin{equation}
\WT(\func{dec(k,d,\gamma)}) = \sum\limits_{j = 1}^{k}{d(u_{j-1},u_{j})} \leq d (1 + \frac{1}{\gamma}) \sum\limits_{j = 1}^{k}{\frac{1}{\gamma^{j-1}}} \leq d k (1 + \frac{1}{\gamma})\nonumber
\end{equation}
And for lower bound on $\WT(\func{dec(k,d,\beta)})$,
\begin{equation}
\WT(\func{dec(k,d,\gamma)}) = \sum\limits_{j = 1}^{k}{d(u_{j-1},u_{j})} \geq d (1 - \frac{1}{\gamma}) \sum\limits_{j = 1}^{k}{\frac{1}{\gamma^{j-1}}} \geq d (1  - \frac{1}{\gamma})\nonumber
\end{equation}
\qed
\end{proof}
Now, for a path $P_{st}$ such that $t$ is a leaf vertex of the tree $T_s$, $P_{st}$ can be written as $\func{inc(r_0,d_0,\beta)} \circ \func{dec(r_1,d_1,\gamma)} \circ \ldots \circ \func{inc(r_{l-1},d_{l-1},\beta)} \circ \func{dec(r_l,d_l,\gamma)}$ (where $\circ$ denotes sequential composition), such that $d_0 = d(s,t)$, $r_l \neq 0$, and for each $i = 1, \ldots, l$ we have $d_i \leq \beta^{r_{i-1}} d_{i-1}$ when $i$ is odd and $d_i \geq d_{i-1}/\gamma^{r_{i-1}}$ when $i$ is even. In other words, $P_{st}$ is a combination of increasing and decreasing sequences with at least one increasing sequence and a trailing decreasing sequence. Also every sequence starts at a distance from $t$, where the immediate previous sequence ends.
\begin{lemma}
\label{L6}
Let $P(k,\beta)$ be a $k$ length $\beta$--weak greedy $st$--path such that $t$ is a leaf vertex of the tree $T_s$. Then 
\begin{equation}
d_{\min}(G) \cdot k \cdot \left(\beta - 1\right) \leq \WT(P(k,\beta)) \leq 2 \cdot d_{\max}(G) \cdot \left(\frac{\beta^k - 1}{\beta-1}\right)\nonumber
\end{equation}
\end{lemma}
\begin{proof}
Let $P$ be composed of $\func{inc(r_0,d_0,\beta)} \circ \func{dec(r_1,d_1,\gamma)} \circ \ldots \circ \func{inc(r_{l-1},d_{l-1},\beta)} \circ \func{dec(r_l,d_l,\gamma)}$. We consider $0$ is even. Using upper bounds on $\WT(\func{inc(k,d,\gamma)})$ and $\WT(\func{dec(k,d,\gamma)})$ from Lemma-\ref{L4} and Lemma-\ref{L5} respectively - length of this sequence is bounded by:
\begin{equation}
\label{EQ1}
d(s,t) (\beta^{r_0} - 1) \left(\frac{\beta + 1}{\beta - 1}\right) + d(s,t) \beta^{r_0} r_1 (1+ \frac{1}{\gamma}) + \ldots + d(s,t) \frac{\beta^{(\sum\limits_{j \in [l-1]: j \text{ even}}{r_j})}}{\gamma^{(\sum\limits_{j \in [l-1]: j \text{ odd}}{r_j})}} r_l (1+ \frac{1}{\gamma}) 
\end{equation}
Or the $i$ the term of this sum can be written as,
\begin{equation}
d_{\func{even}}(i) \define d(s,t) \frac{\beta^{(\sum\limits_{j \in [i-1]: j \text{ even}}{r_j})}}{\gamma^{(\sum\limits_{j \in [i-1]: j \text{ odd}}{r_j})}} (\beta^{r_i} - 1) \left(\frac{\beta + 1}{\beta - 1}\right) \text{ When } i \text{ is even}\nonumber 
\end{equation}
and,
\begin{equation}
d_{\func{odd}}(i) \define d(s,t) \frac{\beta^{(\sum\limits_{j \in [i-1]: j \text{ even}}{r_j})}}{\gamma^{(\sum\limits_{j \in [i-1]: j \text{ odd}}{r_j})}} r_i (1+ \frac{1}{\gamma}) \text{ When } i \text{ is odd}\nonumber
\end{equation}
With constraint that $\sum_{i = 0}^{l}{r_i} = k$, $r_l \neq 0$ and $l$ is odd (since $P$ is $\beta$-weak it can not have only a decreasing sequence, and terminating sequence must be decreasing as $t$ is a leaf vertex). For $d_0 = d(s,t)$ and $k$ fixed, second constraint implies that though sum increases if $\sum_{j \in [i-1]: j \text{ even}}{r_j}$ is maximized and $\gamma$ is close to $1$, this can not be done without increasing $r_l$ and hence decreasing $\sum_{j \in [i-1]: j \text{ even}}{r_j}$. So the expression is maximized with $r_0 = k-1$ and $\gamma = d \beta^{k-1}$. With this we have from equation-\ref{EQ1}:
\begin{eqnarray}
\WT(P(k,\beta)) &\leq& d(s,t) \cdot \left(\beta^{k-1} - 1\right) \cdot \left(\frac{\beta + 1}{\beta - 1}\right) + d(s,t) \beta^{k-1} \left(1 + \frac{1}{\beta^{k-1}}\right)\nonumber\\
&=& 2 \cdot d(s,t) \cdot \left(\frac{\beta^k - 1}{\beta-1}\right) \leq 2 \cdot d_{\max}(G) \cdot \left(\frac{\beta^k - 1}{\beta-1}\right) \nonumber
\end{eqnarray}
Now for the lower bound we consider lower bounds obtained on $\WT(\func{inc(k,d,\gamma)})$ and $\WT(\func{dec(k,d,\gamma)})$ from Lemma-\ref{L4} and Lemma-\ref{L5} respectively. Then we have the length of $P$ lower bounded by:
\begin{equation}
\label{EQ2}
d(s,t) (\beta^{r_0} - 1) + d(s,t) \beta^{r_0} (1 - \frac{1}{\gamma}) + \ldots + d(s,t) \frac{\beta^{(\sum\limits_{j \in [l-1]: j \text{ even}}{r_j})}}{\gamma^{(\sum\limits_{j \in [l-1]: j \text{ odd}}{r_j})}} (1 - \frac{1}{\gamma}) 
\end{equation}
Using equation-\ref{EQ2} with $l = k$, for each $i = 0, \ldots, k-1: r_i = 1$, and $\gamma  = \beta$, we obtain:
\begin{eqnarray}
\WT(P(k,\beta)) &\geq& d(s,t) \cdot k \cdot \left(\beta - 1\right) \geq  d_{\min}(G) \cdot k \cdot \left(\beta - 1\right)\nonumber
\end{eqnarray}
Where, the last inequality follows by taking minimum edge length in embedding of $G$ as $d_{\min}(G)$.
\qed
\end{proof}
Finally we bound the weight of $\beta$-weak greedy spanning tree $T_s$.
\begin{lemma}
\label{L7}
\begin{equation}
d_{\min}(G) \cdot \left(\beta_{\max} - 1\right) \cdot \left(\card{V}-1\right) \leq \WT(T_s) \leq 2 \cdot d_{\max}(G) \cdot \left(\frac{\beta_{max}^{\card{V}-1} - 1}{\beta_{max}-1}\right)\nonumber
\end{equation}
\end{lemma}
\begin{proof}
Assume that $T_s$ has $l$ many leaf nodes. Then weight of the tree is 
\begin{equation}
\WT(T_s) = \sum_{i = 1}^{l}{\WT(P(k_i,\beta))}.\nonumber
\end{equation}
Where $\sum_{i = 1}^{l}{k_i} = \card{V}-1$. In order to obtain the upper bound we observe that $\WT(P(k_i,\beta))$ is maximized with any one of $k_i = \card{V}-1$. Hence using upper bound on $\WT(P(k,\beta))$ from Lemma-\ref{L6} we have: $\WT(T_s) \leq 2 \cdot d_{\max}(G) \cdot (\beta_{max}^{\card{V}-1} - 1)/(\beta_{max} - 1)$. On the other hand, for the lower bound we have $l = \card{V}-1$ and $1 \leq i \leq \card{V}-1 : k_i = 1$. Using lower bound on $\WT(P(k,\beta))$ from Lemma-\ref{L6} we have: $\WT(T_s) \geq d_{\min}(G) \cdot \left(\beta_{\max} - 1\right) \cdot \left(\card{V}-1\right)$
\qed
\end{proof}
%
%
\subsection{Bound on $\beta_{\max}$}
\label{SEC45}
As stated in the beginning of this section, we now compare the bound on the weight of any spanning tree $T$ of $G$ with that of $T_s$ as derived in Lemma-\ref{L3}, Lemma-\ref{L8} and Lemma-\ref{L7} to obtain an upper and lower bound on $\beta_{\max}$.
\begin{theorem}
\label{THM:T3}
Let $G = (V,E)$ be any three connected planar graph. Then $G$ has a $\beta$-weak greedy convex embedding in $\bbbr^2$ with 
\begin{equation}
\beta \in [1,2\sqrt{2} \cdot d(G)].\nonumber
\end{equation}
Also, this bound is achieved by Tutte embedding.
\end{theorem}
\begin{proof}
Let $T_s$ be any $\beta$-weak greedy spanning tree of $G$ with respect to vertex $s \in V$. Let $T$ be any spanning tree of $G$, and let $\func{MST}$ be any minimum weight spanning tree of $G$. Then using Lemma-\ref{L8}, and upper bound on the $\WT(T_s)$ from Lemma-\ref{L7} we obtain:
\begin{eqnarray}
\WT(T_s) &\geq& \WT(\func{MST})\nonumber\\
2 \cdot d_{\max}(G) \cdot \left(\frac{\beta_{max}^{\card{V}-1} - 1}{\beta_{max}-1}\right) \geq \WT(T_s) &\geq& \WT(\func{MST}) \geq \WT(\func{EMST}) \geq d_{\max}(G) \nonumber
\end{eqnarray}
Which implies:
\begin{eqnarray}
\left(\frac{\beta_{max}^{\card{V}-1} - 1}{\beta_{max}-1}\right) &\geq& \frac{1}{2} 
\label{T3:EQ1}
\end{eqnarray}
And this holds for any $\beta_{max} > 1$ when $\card{V} \geq 3$. On the other hand using Lemma-\ref{L3}, and lower bound on the $\WT(T_s)$ from Lemma-\ref{L7}:
\begin{eqnarray}
\WT(T_s) &\leq& \WT(T)\nonumber\\
d_{\min}(G) \cdot \left(\beta_{\max} - 1\right) \cdot \left(\card{V}-1\right) \leq \WT(T_s) &\leq& \WT(T) \leq \sqrt{2} \cdot (\card{V} - 1) \cdot d_{\max}(G)\nonumber
\end{eqnarray}
Now using $d(G) = d_{\max}(G)/d_{\min}(G)$ we have:
\begin{eqnarray}
\beta_{max} &\leq& \sqrt{2} \cdot \frac{d_{\max}(G)}{d_{\min}(G)} + 1 \leq \sqrt{2} \cdot d(G) + 1 \leq 2\sqrt{2} \cdot d(G)
\label{T3:EQ2}
\end{eqnarray}
Finally, to show that this bound is tight consider Tutte embedding of a cube (see figure-\ref{fig6A}) with all edges assigned with same weights. It can be seen that in this embedding $\beta \leq 1$. On the other hand, when we reduce the weight on the edges $BF$ and $DH$ (see figure-\ref{fig6B}) we obtain an embedding in which there is no greedy path between pair $B$ and $D$, while there is a $\beta$-weak greedy path with $\beta$ approaching $d(G)/2$.
\begin{figure*}[htbp]
\centering
\subfigure[Equal edge weights]{\includegraphics[viewport=100 100 230 220,width=0.35\textwidth,clip]{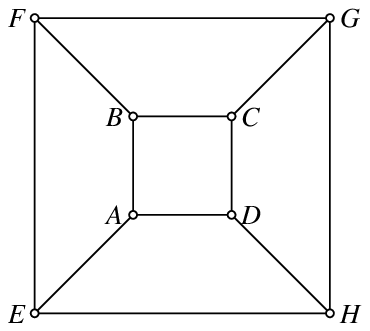}\label{fig6A}}
\subfigure[Unequal edge weights]{\includegraphics[viewport=100 100 230 220,width=0.35\textwidth,clip]{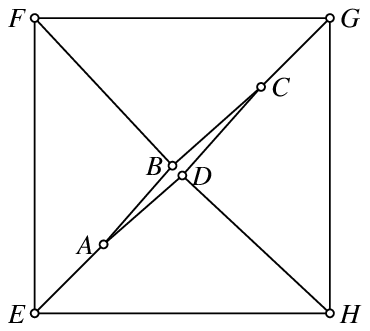}\label{fig6B}}
\caption{Illustration of Tutte embedding of a cube}
\label{fig6}
\end{figure*}
\qed
\end{proof}
If we consider Tutte embedding of a $3$--connected planar graph $G$ with arbitrary weights on the edges, then it is not difficult to see that above bound on $\beta$ depends entirely on the choice of the edge weights in the Tutte embedding. 
\section{Characterizing convex greedy embedding}
\label{SEC5}
\begin{theorem}
\label{THM:T1}
For sufficiently large $\card{V}$ for a $3$-connected planar graph $G = (V,E)$ if embedding $x: V \rightarrow \bbbr^2$ of $G$ is such that the maximum weight spanning tree ($T$) and minimum weight spanning tree ($\func{MST}$) satisfies: 
\begin{equation}
\frac{\WT(T)}{\WT(\func{MST})} \leq \left(\card{V}-1\right)^{1 - \delta}, \text{ for some } 0 < \delta \leq 1.\nonumber
\end{equation}
Then embedding $x$ is a convex greedy embedding of $G$.
\end{theorem}
\begin{proof}
Observe that we have following relations:
\begin{equation}
\WT(\func{MST}) \leq \WT(T_s) \leq \WT(T) \nonumber
\end{equation}
Since $\WT(\func{MST}) > 0$, using lower bound on $\WT(T_s)$ from Lemma-\ref{L7} and using upper bound on $\WT(\func{MST})$ from Lemma-\ref{LMSTUB} we obtain:
\begin{equation}
\frac{2 \cdot d_{\min}(G) \cdot \left(\beta_{\max} - 1\right) \cdot \left(\card{V}-1\right)}{5 \cdot d^2_{\max}(G)} \leq  \frac{\WT(T)}{\WT(\func{MST})} \nonumber
\end{equation}
\begin{equation}
\text{And hence, } \beta_{\max} \leq \left(\frac{5 \cdot d_{\max}(G) \cdot d(G)}{2 \cdot \left(\card{V}-1\right) } \right) \cdot \left(\frac{\WT(T)}{\WT(\func{MST})}\right) + 1 \nonumber
\end{equation}
Now if weight of the maximum and minimum spanning tree in the planar convex embedding of $G$ is such that ${\WT(T)}/{\WT(\func{MST})} \leq \left(\card{V}-1\right)^{1 - \delta}$ for some $0 < \delta \leq 1$, then for sufficiently large $\card{V}$, $\beta_{\max} \rightarrow 1$ from above (note that $\beta_{\max} > 1$ by Equation-\ref{T3:EQ1}).
\qed
\end{proof}
In following we show the more interesting direction:
\begin{theorem}
\label{THM:T2}
Given a $3$-connected planar graph $G = (V,E)$, if embedding $x: V \rightarrow \bbbr^2$ of $G$ is a convex greedy embedding then in embedding $x$ the maximum weight spanning tree ($T$) and minimum weight spanning tree ($\func{MST}$) satisfies: 
\begin{equation}
\frac{\WT(T)}{\WT(\func{MST})} \leq \left(\card{V}-1\right)^{1 - \delta}, \text{ for some } 0 < \delta \leq 1.\nonumber
\end{equation}
\end{theorem}
\begin{proof}
For a $3$-connected planar graph $G = (V,E)$, let an embedding $x: V \rightarrow \bbbr^2$ of $G$ be a convex greedy embedding. Let us also assume that ${\WT(T)}/{\WT(\func{MST})} \geq \left(\card{V}-1\right)$. W.l.o.g. let $\WT(\func{MST}) = 1$. Since $T$ is a spanning tree it has $\left(\card{V}-1\right)$ edges, and hence has at least one edge $e \in T$ of weight $\WT(e) \geq 1$. Given that $x$ is a convex planar embedding of a $3$-connected planar graph $G$, we have that each edge belongs to exactly two faces of the graph (in fact a graph is $3$-connected and planar if and only if each edge is in exactly two non-separating induced cycles \cite{Kelmans78}). So we consider two cases: (Case - 1) $e$ is on two internal faces $F$ and $F'$, and (Case - 2) $e$ is on the boundary face. We need few  definitions \cite{Kelmans00}. For a graph $G$, a \emph{thread} is a path $P$ of $G$ such that any degree $2$ vertex $x$ of $G$ is not an end vertex of $P$. A sequence $S = (G_0, \set{x_i P_i y_i : i = 1, \ldots,k})$ is an \emph{ear-decomposition} of $G$ if:
\begin{enumerate}
	\item $G_0$ is a subdivision of $K_4$,
	\item $x_i P_i y_i$ is a path with end-vertices $x_i$ and $y_i$ such that $G_i = G_{i-1} \cup P_i$ is a subgraph of $G$, and $G_{i-1} \cap P_i = \set{x_i,y_i}$, but $x_i$, $y_i$ do not belong to a common thread of $G_{i-1}$ for $i = 1,\ldots, k$, and 
	\item $G_k = G$.
\end{enumerate}
We will need following result from \cite{Kelmans00}:
\begin{lemma}[\cite{Kelmans00}]
\label{LEM:Kelmans00}
Let $G$ be a $3$--connected graph, $e = uv \in E(G)$. Let $C_1$ and $C_2$ be non-separating cycles of $G$ such that $C_1 \cap C_2 = uev$. Then there exists an ear-decomposition of $G$ such that $C_1 \cup C_2 \subset G_0$.
\end{lemma}
\paragraph{Case - 1:} In this case $e = uv$ is on two internal faces $F_1$ and $F_2$. Consider a vertex $u'$ from face $F_1$ and another vertex $v'$ from face $F_2$. First consider $K_4$, which has four faces, and exactly one planar convex embedding. However, vertices $u, v, u', v'$ must be spanned by the $\func{MST}$ using exactly $3$ edges. If $e$ is chosen in the $\func{MST}$ then other edges are of length $0$, as $\WT(e) \geq 1$ and $\WT(\func{MST}) = 1$. If $e$ is not selected in $\func{MST}$ - then it can be easily seen that either $\WT(\func{MST}) > 1$, or the drawing is not planar - a contradiction. In specific this can be seen as follows (see Figure-\ref{fig7}): consider that $uu'$,$u'v$ and $u'v'$ is selected in $\func{MST}$ - then we have $uu' + u'v \geq uv$ (where, $uv$ is an edge in the external face $uvu'$) and this implies either $uu' + u'v + u'v' > uv \geq 1$, or $u'v' = 0$. Now, let $G$ be a $3$-connected planar graph that is distinct from $K_4$. Then there exists an ear-decomposition of $G$ such that $e = uv$ and faces $F_1$ and $F_2$ are such that $F_1 \cup F_2 \subset G_0$, where $G_0$ is a subdivision of $K_4$, by Lemma-\ref{LEM:Kelmans00}. We can contract edges of $F_1 \cup F_2$ while keeping edge $e$ to obtain a $K_4$. In this process we never increase the weight of the $\func{MST}$, and hence obtain the contradiction as above.
\begin{figure}[htbp]
\centering
\includegraphics[viewport=0 0 300 120,width=0.65\textwidth,clip]{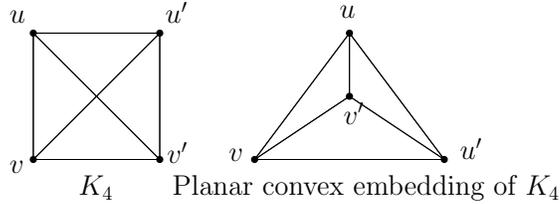}
\caption{Illustration to the proof of Case - 1 for $K_4$}
\label{fig7}
\end{figure}
\paragraph{Case - 2:}In this case $e = uv$ is on the boundary face. Since minimal external face must be a triangle there exists another vertex $u'$ on the external face. Consider another internal vertex $v'$. Again vertices $u, v, u', v'$ must be spanned by the $\func{MST}$ using at least $3$ edges. If $e$ is chosen in the $\func{MST}$ then other edges are of length $0$, as $\WT(e) \geq 1$ and $\WT(\func{MST}) = 1$. On the other hand if $e$ is not selected in $\func{MST}$ - then $\WT(\func{MST}) > 1$ if embedding is convex, a contradiction.
\qed
\end{proof}
%
%
\section{Concluding remarks}
\label{SEC6}
With Theorem-\ref{THM:T1} and Theorem-\ref{THM:T2}, thus, with the example presented above (Figure-\ref{fig6}) we can ask following question: For every $3$--connected planar graph $G$, is it possible to choose edge weights in the Tutte embedding such that we obtain a greedy convex embedding? We believe that answer to this question will help in making progress towards resolving original convex greedy embedding conjecture of Papadimitriou and Ratajczak \cite{PR05}.\par
We would like to clarify that though the $\beta$--weak greedy path finding algorithm presented above is stateless, it is not a practical routing algorithm - as number of messages will be large even for constant values of $\beta$, when $\beta > 1$, and the routing procedure also forms cycles. The purpose of defining $\beta$--weak greedy path finding procedure was to derive the main results of this paper.
%
\bibliographystyle{splncs}
\bibliography{ge3cpg}  
\appendix
\section{Tutte Embedding}
\label{APX2}
\begin{figure*}[htbp]
\centering
\subfigure[Embedding using a face having $12$ vertices]{\includegraphics[viewport=0 0 510 510,width=0.25\textwidth,clip]{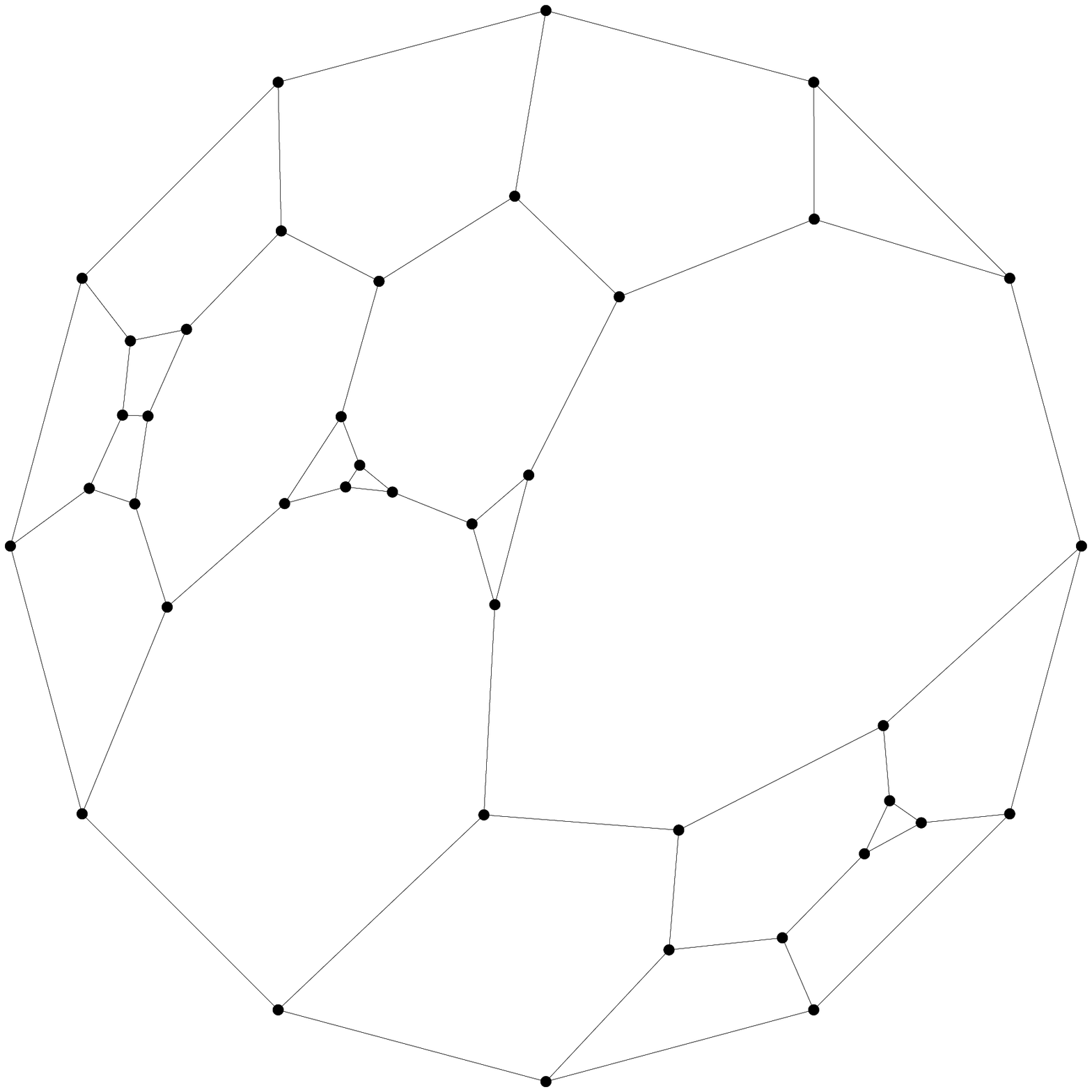}\label{fig1A}}
\subfigure[Another embedding of same graph]{\includegraphics[viewport=0 0 510 510,width=0.25\textwidth,clip]{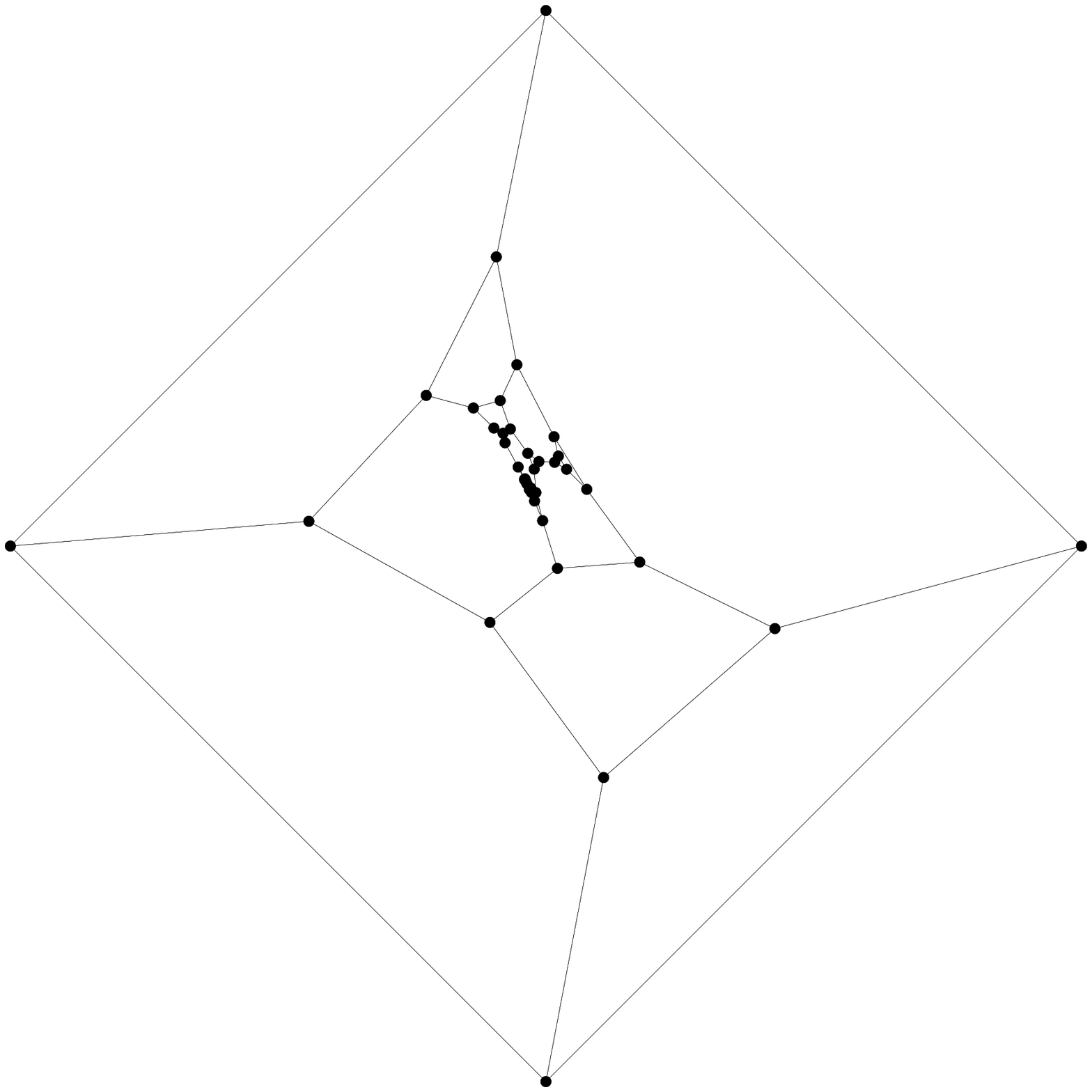}\label{fig1B}}
\subfigure[$\beta$--weak greedy path between $s$ and $t$]{\includegraphics[viewport=0 0 510 510,width=0.25\textwidth,clip]{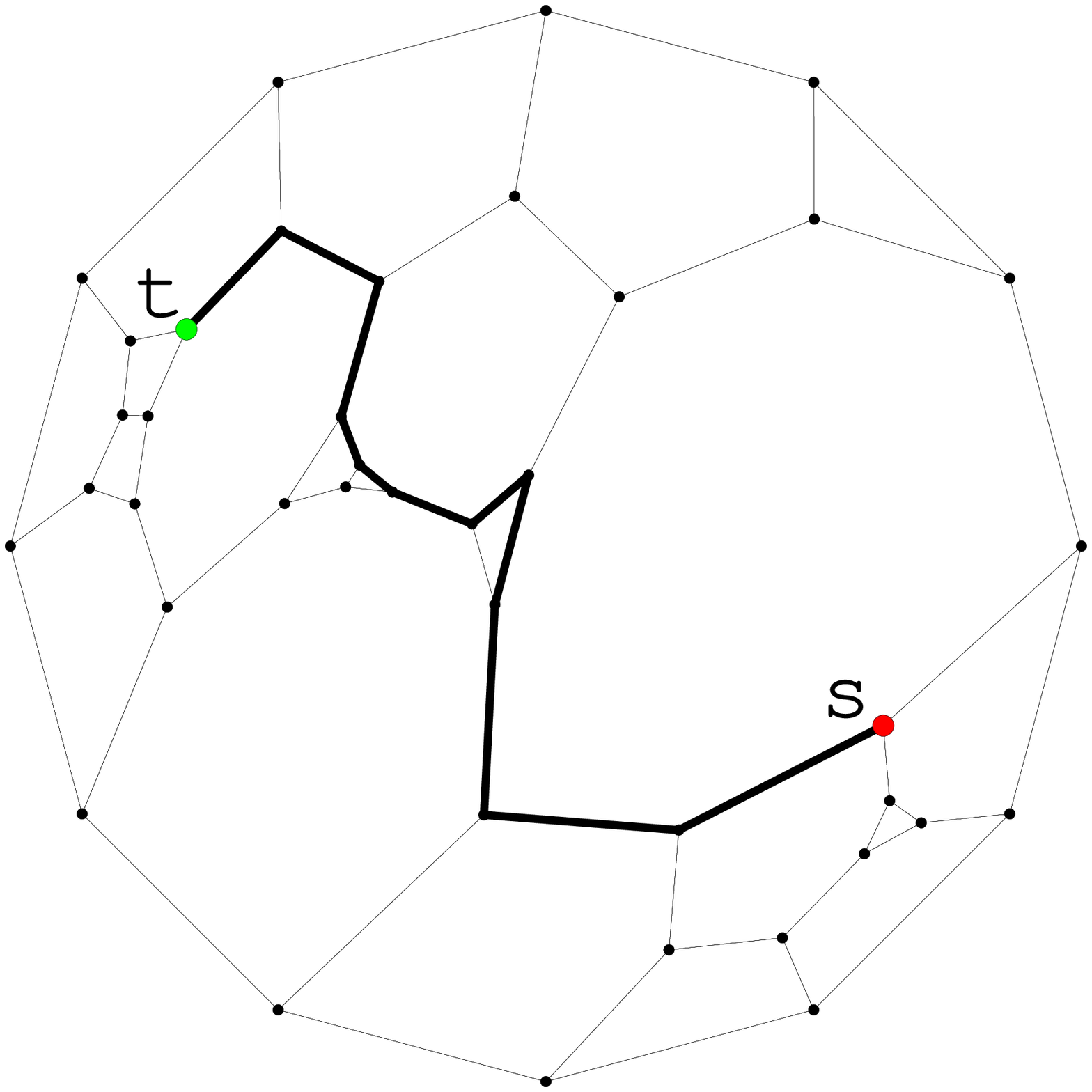}\label{fig1C}}
\subfigure[Corresponding distance function]{\includegraphics[viewport=0 0 400 370,width=0.25\textwidth,clip]{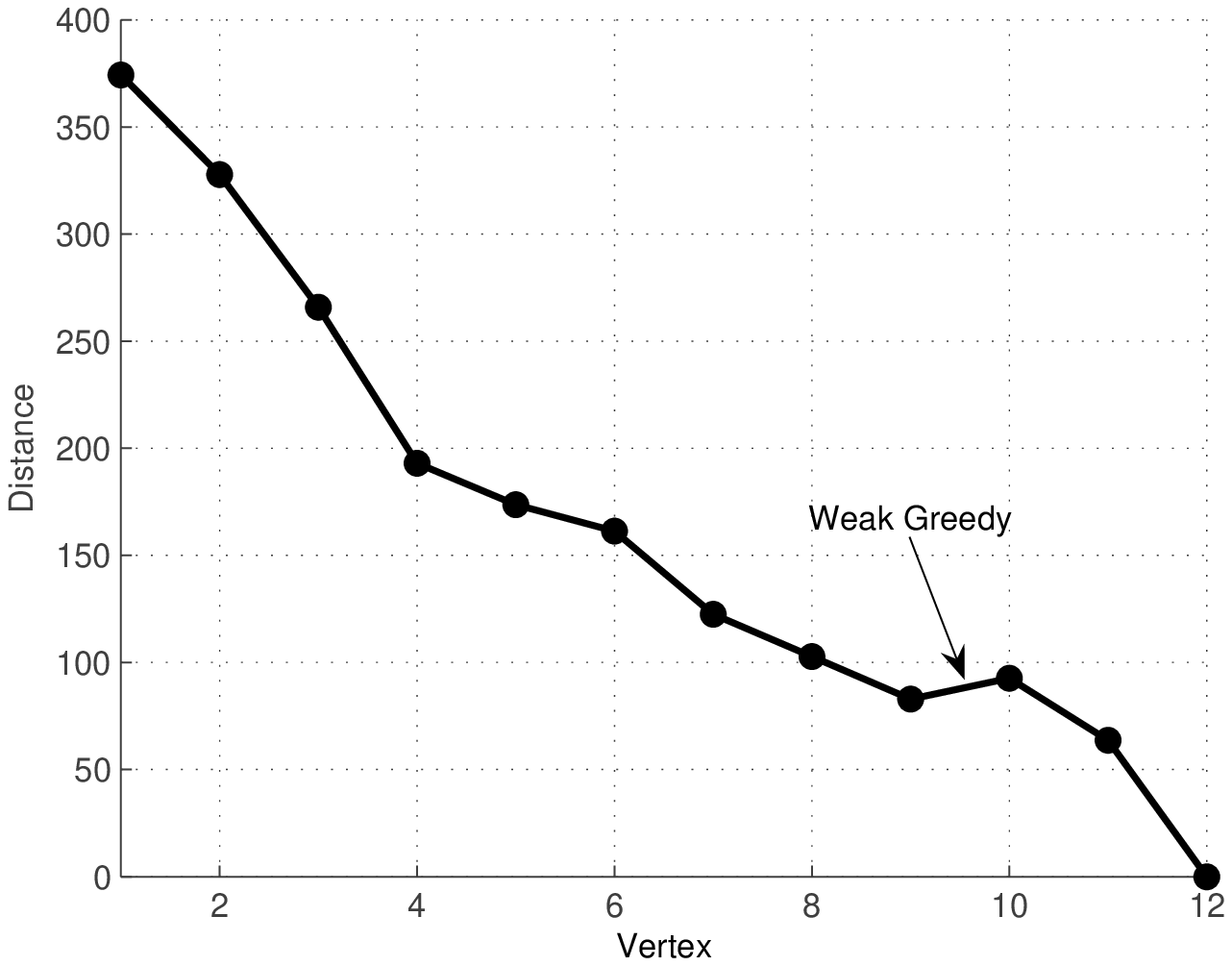}\label{fig1D}}
\caption{Illustration of Tutte embedding of a $3$--connected planar graph, and weak greedy routing}
\label{fig1}
\end{figure*}

We mentioned that every $3$-connected planar graph has a convex embedding in the the Euclidean plane using Tutte's rubber band algorithm \cite{Tutte}. Here we provide a short description of this. Let $G = (V,E)$ be a $3$--connected planar graph and $\emptyset \neq S \subseteq V$. Let $x^0 : S \rightarrow \bbbr^2$ be a map. We extend $x^0$ to a geometric representation of $G$, $x : V \rightarrow \bbbr^2$ as follows. We consider each edge $uv \in E$ is made of ideal rubber band that follows Hook's law and is assigned with a positive weight $w_{uv}$, and each node $u \in S$ has a \emph{nailed} position as given by $x^0(u) \in \bbbr^2$. Other nodes $v \in V \setminus S$ then come to an equilibrium. For a node $u \in V$, let $x(u) \in \bbbr^2$ be its position. The \emph{energy} of this representation is defined by function 
\begin{equation}
{\mathcal E}(x) = \frac{1}{2} \sum\limits_{uv \in E}{w_{uv} \cdot d^2\left(u,v\right)}.\nonumber
\end{equation}
At the equilibrium, ${\mathcal E}(x)$ is minimized subject to the boundary conditions namely, nailed positions of the vertices in $S$. First note that ${\mathcal E}(x)$ is strictly convex as $d^2(\cdot,\cdot)$ is whenever $S \neq \emptyset$. Also there is a unique optimum and in optimal representation 
\begin{equation}
\forall u \in V \setminus S: \sum\limits_{v \in {\mathcal N}(u)}{w_{uv} \cdot \left(x(u) - x(v)\right)} = 0\nonumber
\end{equation}
Or, every vertex $ u \in V \setminus S$ is in the relative interior of the convex hull of its neighbors as 
\begin{equation}
\forall u \in V \setminus S: x(u)  = \frac{1}{\sum\limits_{v \in {\mathcal N}(u)}{w_{uv}}} \cdot \sum\limits_{v \in {\mathcal N}(u)}{w_{uv} \cdot x(v)}\nonumber
\end{equation}
Tutte's result states that:
\begin{theorem}[\cite{Tutte}]
\label{THM:TUTTE}
Let $G = (V,E)$ be a $3$--connected planar graph , $F$ be any face of $G$ and $C$ be cycle bounding $F$ (call it external face). Define $w : E \setminus E(C) \rightarrow \bbbr^{+}$, and $x^0: V(C) \rightarrow \bbbr^2$. Then:
\begin{enumerate}
	\item $x^0$ extends to $x: V \rightarrow \bbbr^2$ such that all vertices $u \in V \setminus V(C)$ has unique representation $x(u) \in \bbbr^2$ when in equilibrium.
	\item Boundary of every internal face of $G$ is realized as convex polygons such that their interiors are disjoint.
\end{enumerate}
\end{theorem}
We shall further assume that if external face has $k$ vertices, then $x^0$ maps them (maintaining the order of the cycle) to a $k$-gon in $\bbbr^2$. There are several exposition of the proof of Theorem-\ref{THM:TUTTE} and we suggest interested reader to refer \cite{GR01,Ric96}. Note that the embedding itself is not unique, and it depends on the choice of the external face (e.g. see Figure-\ref{fig1A} and \ref{fig1B}).
\end{document}